\documentstyle[aps,eqsecnum]{revtex}

\def\be{\begin{equation}}
\def\ee{\end{equation}}
\def\ba{\begin{eqnarray}}
\def\ea{\end{eqnarray}}

\begin{document}


\draft
\title{Do stringy corrections stabilize coloured
black holes?}

\author{P. Kanti}
\address{Theoretical Physics Institute, School of Physics
and Astronomy, University of Minnesota, Minneapolis,
MN 55455, USA.}

\author{E. Winstanley}
\address{Department of Physics (Theoretical Physics),
1, Keble Road, Oxford, OX1 3NP, UK.}

\date{\today}
\maketitle

\begin{abstract}
We consider hairy black hole solutions of Einstein-Yang-Mills-Dilaton
theory, coupled to a Gauss-Bonnet curvature term, and we study their
stability under small, spacetime-dependent perturbations.
We demonstrate that the stringy corrections do not remove the sphaleronic
instabilities of the coloured black holes with the number of unstable
modes being equal to the number of nodes of the background gauge
function. In the gravitational sector, and in the limit of
an infinitely large horizon, the coloured black holes are also
found to be unstable. Similar behaviour is exhibited by the
magnetically charged black holes while the bulk of the neutral
black holes are proven to be stable under small, gauge-dependent
perturbations.
Finally, the electrically charged black holes are found to be
characterized only by the existence of a gravitational sector of
perturbations. As in the case of neutral black holes, we
demonstrate that for the bulk of electrically charged black holes
no unstable modes arise in this sector. 
\end{abstract}

\pacs{PACS numbers: 04.70.Bw, 04.20.Jb, 04.50.+h, 11.25.Mj\\
UMN-TH-1823/99; TPI-MINN-99/48; OUTP-99-51-P; gr-qc/9910069}

\section{Introduction}
To date a large number of hairy black hole geometries have been 
found, in a diverse range of matter models coupled to various theories
of gravity (for a recent review of some aspects, see \cite{review}).  
Stable examples of black holes with hair are of particular interest
for their physical relevance, and also in order to investigate 
the effects of hair on quantum processes connected with black holes.
However, many of the hairy black holes currently known
are unstable, particularly those involving non-Abelian gauge
fields~\cite{strau,lavr,don,maeda,volcount,em,higgs}, where the instability
is topological in nature and similar to that of the flat space sphaleron.
The only exceptions to the above rule are the black hole solutions found
in the framework of the Einstein-Skyrme theory~\cite{skyrme} and the
magnetically charged, non-Abelian black holes in the limit of infinitely
strong coupling of the Higgs field~\cite{ab} or in the presence of a
negative cosmological constant~\cite{ew}. The limited number of stable
black holes known so far makes the quest for new stable solutions both
immediate and challenging.

The superstring effective action at low energies, that follows from the
compactification of the heterotic superstring theory, provides us with
a generalized theory of gravity which is an excellent framework for the
study of black holes. The theory contains, apart from the usual Einstein
term, a number of scalar fields, the dilaton, axion and moduli fields,
coupled to higher-derivative gravitational terms as well as non-Abelian
gauge fields. Here, we are particularly interested in the theory that
describes the coupling of the dilaton field to the one-loop Gauss-Bonnet
curvature term, and can also include a non-Abelian gauge field.
These models are known to possess black hole solutions with hair, both
with \cite{kt,torii,sa} and without \cite{kmrtw1,sps} the gauge field.
In the absence of the gauge field, the so-called dilatonic hairy black
holes found in \cite{kmrtw1} were proven to be linearly stable \cite{kmrtw2}
with their stability being in accordance with their interpretation as a
generalization of the Schwarzschild black hole in the framework of
string theory. This family of stable black holes corresponds to the
upper branch of the neutral black hole solutions found later in
\cite{torii} and whose relative stability with respect to a second
unstable branch of solutions was demonstrated by the use of catastrophe
theory~\cite{extra}.
However, the question of the stability of the corresponding
coloured black holes, that arise in the presence of
the non-Abelian gauge field, still remains open. In this article we
investigate whether the stability of the dilatonic black holes
extends to the case where the dilaton is also coupled to the gauge field. 
We will be particularly interested in stringy coloured black holes,
and it will be shown that they, like their non-stringy counterparts,
are topologically unstable. In addition, conclusions on the stability
of the dilatonic black holes under small, gauge perturbations as well
as that of the magnetically and electrically charged black holes will
also be drawn. 

The outline of the paper is as follows. 
In section II we introduce our model and briefly review
the properties of the stringy black hole solutions.
The solutions fall into four categories: neutral, coloured, 
magnetically charged and electrically charged, depending on 
the behaviour of the gauge field. For the first three types of solutions,
the linearized perturbation equations decouple, under an appropriate
choice of gauge, into two sectors, corresponding to
gravitational and sphaleronic perturbations. 
We first concentrate on the sphaleronic sector,
and show in section III that there are topological instabilities
for both coloured and magnetically charged black holes.
We then count the number of unstable modes in this sector and
find that it equals the number of zeros of the gauge field function
of the background solution. In the same section, we also consider
the stability of the sphaleronic sector of the neutral black holes.
In section IV, we comment on the gravitational sector
for all four types of black holes, appealing to catastrophe theory and
continuity arguments.
The magnetically charged and coloured black holes have instabilities
in this sector due to the presence of the non-Abelian gauge field,
whilst the bulk of the electrically charged and neutral black holes
are shown to be stable. Section V is devoted to the conclusions
derived from our analysis.
 
\section{Stringy coloured black holes}

We consider the following Lagrangian describing  
strin\-gy corrections to the SU(2) Einstein-Yang-Mills model
\cite{kt}
\be
{\cal {L}}=
\frac{R}{2} + \frac{1}{4}\,
\partial_\mu\phi\,\partial^\mu\phi + \frac{\alpha'e^\phi}{8g^2}\,
(\beta {\cal R}^2_{GB}
-F^{a\mu\nu} F^a_{\,\,\mu\nu}) \,\,,
\label{action}
\ee
where $\phi$ is the dilaton field and ${\cal R}^2_{GB}=R_{\mu\nu\rho\sigma}
R^{\mu\nu\rho\sigma}-4R_{\mu\nu} R^{\mu\nu} + R^2$ the higher-derivative
Gauss-Bonnet term.
In equation (\ref{action})  we have set the gravitational coupling
$\kappa ^{2}=8\pi G$ equal to unity, so that 
$\alpha '/g^{2}$ is the single coupling constant.
In effective string theory, which is the situation 
in which we are interested, the constant $\beta $ is 
equal to unity, and from now on we shall fix
$\beta $ to have this value.
It is known that, in this case, there are no particle-like 
solutions to the field equations \cite{donets}.
In this paper, we are concerned only with the black hole solutions, 
and shall not consider the particle-like solutions 
which exist for other values of $\beta $.
However, much of our analysis applies equally well for all values of
$\beta $, and in particular to the case $\beta =0$, which 
corresponds to the Einstein-Yang-Mills-Dilaton (EYMD)
black holes considered in \cite{lavr,don}.
If the coupling constant is fixed (for numerical computations the 
value $\alpha '/g^{2}=1$ is convenient) then there is only one 
remaining parameter, namely the event horizon radius $r_{h}$.

We consider the general spherically symmetric ansatz for
the $SU(2)$ non-Abelian gauge potential \cite{witten}
\ba
{\bf A}& = & a_0 \hat{\tau}_r\,dt + b \hat{\tau}_r\,dr +
[\nu \hat{\tau}_\theta - (1+w) \hat{\tau}_\varphi]\,d\theta +
[(1+w) \hat{\tau}_\theta + \nu \hat{\tau}_\varphi]\,
\sin \theta\,d\varphi\,\,,
\label{ansatz}
\ea
with 
${\hat {\tau }}_{r}={\mbox {\boldmath {${\hat {\tau }}.e_{r}$}}}$,
where ${\hat {\tau }}_{i}$, $i=1,2,3$ are the usual Pauli matrices.
This ansatz for the gauge potential (\ref{ansatz}) does not completely
fix the gauge.  
There is still freedom to make unitary transformations of the
form
\be
{\bf {A}} \rightarrow {\cal {T}} {\bf {A}} {\cal {T}}^{-1}
+{\cal {T}} d{\cal {T}}^{-1}
\ee
where
\be
{\cal {T}}=\exp \left[ k(r,t) {\hat {\tau }}_{r} \right] ,
\ee 
under which the gauge potential components transform according to
\be
\left(
\begin{array}{c}
a_{0} \\ b \\ w \\ \nu 
\end{array}
\right) \rightarrow \left(
\begin{array}{c}
a_{0}-{\dot {k}} \\
b-k' \\
w \cos k -\nu \sin k \\
\nu \cos k +w \sin k 
\end{array}
\right) ,
\label{gaugefree}
\ee
where here, and in the rest of the paper, we use $'$ to denote
derivative with respect to $r$, and ${\dot {}}$ to denote derivative
with respect to $t$.
We shall make use of this gauge freedom to choose the gauge
for which the perturbation equations take their simplest form.

We also consider the following spherically symmetric line-element for
the spacetime background
\be
ds^2= -e^{\Gamma}\,dt^2+e^{\Lambda}\,dr^2 +r^2\,
(d\theta^2+\sin^2\theta\,d\varphi^2)\,\,.
\ee
In the above two ansatzes for the non-Abelian gauge field and the
line-element, $\Gamma $, $\Lambda $, $a_0$, $b$,
$\nu$ and $w$ depend on $t$ and $r$. 
For the equilibrium solutions,  $b$ and $\nu $ vanish identically, 
and all quantities depend on the co-ordinate $r$ only.
Solutions are found by numerical integration, requiring that there
is a regular event horizon at $r=r_{h}$ and that the geometry is 
asymptotically flat, with no singularities outside the event horizon.
The requirement of asymptotic flatness places the following 
constraints on the matter fields as $r\rightarrow \infty $:
the dilaton $\phi  \rightarrow  0$, whilst the gauge field function
$a_0 \rightarrow  0$. 
Finally, $w  \rightarrow \pm 1$ for the coloured black holes with
vanishing charge at infinity, and
$ w \rightarrow 0$ for the globally magnetically charged solutions.
The types of solutions fall into four categories, 
depending on the form of the gauge field:
\begin{enumerate}
\item
{\em {Neutral}} black holes, for which $w\equiv \pm 1$ and 
$a_{0}\equiv 0$.
These solutions have a vanishing gauge field strength
and were discussed in \cite{kmrtw1,kmrtw2}.
\item
{\em {Magnetically charged}} black holes, where $a_{0}\equiv 0$ and 
$w\equiv 0$.  
These have a fixed magnetic charge of unity once the 
coupling constants are fixed \cite{torii}.
\item
{\em {Coloured}} black holes, where $a_{0}\equiv 0$ and $w$ varies.  
These are the stringy sphalerons described in \cite{kt}.
The function $w$ has at least one zero, and the solutions are 
characterized by the number of nodes of $w$.
\item
{\em {Electrically charged}} black holes, where 
$a_{0}$ does not vanish.  
The equations describing these black holes are found from
the general field equations (derived by varying the action
(\ref{action})) by setting $c\equiv a_{0}w\equiv 0$ and then
$w\equiv \pm 1$.
The form of the function $a_{0}(r)$ is fixed by the field 
equations to satisfy
\be
a_{0}'=e^{(\Gamma +\Lambda )/2}e^{-\phi } \frac {Q}{r^{2}},
\ee
where the electric charge $Q$ of the black hole 
can be varied, provided a naked singularity 
is not formed \cite{torii}.
\end{enumerate}
The form of the metric and dilaton fields are qualitatively 
the same for all four types of solutions, with the 
dilaton field monotonically decreasing to its asymptotic value
and the metric functions interpolating between their horizon
and asymptotic values at infinity. Regular black hole solutions
of each type exist for all values of $r_{h}$ above a critical value
at which point a naked singularity is formed. In the case of
dilatonic and coloured black holes with a Gauss-Bonnet term
the above constraint takes the form~\cite{kt,kmrtw1}
\be
\frac{\alpha'}{g^2} e^{\phi_h} < \frac{r_h^2}{\sqrt{6}}\,.
\label{constraint}
\ee
The first three types of solution are easily studied within 
the same algebraic framework, simply choosing $w$ as 
appropriate for each case.
Since the electrically charged black holes have a non-zero 
function $a_{0}$, they will have to be considered separately 
in the stability analysis.

\section{Sphaleronic instabilities}

In this section we consider the magnetically charged and coloured
black holes and investigate the existen\-ce of sphaleronic instabilities
under small, bounded, spacetime-dependent perturbations. 
We use the {\it temporal gauge} and set $a_0=0$. 
The advantage of using the temporal gauge is that the system of 
perturbation equations decouples into two sectors, 
the {\em {gravitational}} sector consisting of $\delta \Gamma $,
$\delta \Lambda $, $\delta \phi$ and $\delta w$, and the 
{\em {sphaleronic}} sector which comprises $\delta b$ and 
$\delta \nu $.
For the electrically charged black holes, the most useful choice
of gauge is not immediately apparent.
This will be considered in section IV, where it is shown
that the electrically charged black holes
effectively have only  gravitational sector perturbations.
Here, we concentrate on the study of the sphaleronic sector of
the coloured and magnetically charged black holes leaving the
stability analysis of their gravitational sector also for section IV.
Our work in this section will also be applicable to the stability
analysis of neutral black holes under small, gauge perturbations, and
the corresponding conclusions will be drawn simply by setting 
$ w \equiv \pm 1 $ in the following. The analysis of subsection IIIB
will then confirm that the neutral black holes do not have
any instabilities in this sector.

For the sphaleronic sector, and the coloured and magnetically 
charged black holes,
we have the following perturbation
equations, where we have considered periodic perturbations
of the form 
$\delta {\cal {P}}(r,t)=\delta {\cal {P}} (r) 
e^{i\sigma t}$:
\be
\sigma \left[
\delta b' + \delta b \,\biggl(\phi' + \frac{2}{r} -
\frac{\Gamma '}{2}-\frac{\Lambda '}{2}\biggr) + 
\frac{2e^\Lambda}{r^2}\,w\,\delta \nu \right]
=0;
\label{2}
\ee
and
\be
2e^{\Gamma -\Lambda }
\left(
\begin{array}{cc}
{\cal {H}}_{bb} & {\cal {H}}_{b\nu } \\
-{\cal {H}}_{\nu b} & -{\cal {H}}_{\nu \nu }
\end{array}
\right) 
 =\sigma ^{2} 
\left(
\begin{array}{cc}
r^{2}e^{-\Lambda } & 0 \\
0 & 2 
\end{array}
\right) \left(
\begin{array}{c}
\delta b \\
\delta \nu 
\end{array} \right) ,
\label{varsyst}
\ee
where
\ba
{\cal {H}}_{bb} & = & 
w^{2} \, \delta b 
\nonumber \\
{\cal {H}}_{b\nu } &  = & 
w \, \delta \nu ' - w' \, \delta \nu 
\nonumber \\
{\cal {H}}_{\nu b}  & = & 
\left( w \, \delta b \right) ' +
w\left( \phi ' +\frac {\Gamma '-\Lambda '}{2} \right) \delta b
+ w' \, \delta b 
\nonumber \\
{\cal {H}}_{\nu \nu } &  = &  
\delta \nu '' + \left( \phi ' +\frac {\Gamma '-\Lambda '}{2} \right) 
\delta \nu '
+\frac {e^{\Lambda }}{r^{2}} (1-w^{2}) \delta \nu  .
\ea
These equations depend on the dilaton field $\phi $, 
however, they have no dependence on the Gauss-Bonnet curvature term, except
through the static solutions. Therefore, our analysis applies equally
well to the case of the Einstein-Yang-Mills-Dilaton (EYMD) black holes.
In the following subsections, we shall first of all prove the existence
of unstable modes for both the magnetically charged and coloured black
holes, and then proceed to count the number of modes of instability.
Our analysis will not depend on the details of the equilibrium 
solutions, only on properties such as the number of nodes of the gauge
field.
In this sense the instabilities are ``topological''.

\subsection{Existence of instabilities}
In this subsection we shall employ a variational method to show that
there are sphaleronic instabilities for both the coloured 
and magnetically charged black holes.
The sphaleronic sector perturbation equations (\ref{varsyst}) 
take the form
\be 
{\cal {H}} \left(
\begin{array}{c} \delta b \\ \delta \nu
\end{array} \right)
=\sigma ^{2} {\cal {A}} \left(
\begin{array}{c} \delta b \\ \delta \nu
\end{array} \right) .
\ee
The operator ${\cal {H}}$ must be self-adjoint with respect to 
a suitable inner product on the space of functions 
$\Psi =\left( \begin{array}{c} \delta b \\ \delta \nu 
\end{array} \right) $, and
the operator ${\cal {A}}$ is required to be positive definite, 
so that $\langle \Psi | {\cal {A}} |\Psi \rangle >0$ 
for all non-zero $\Psi $, using the same inner product.
The variational method has been applied successfully to 
various systems involving non-Abelian gauge fields 
(see, for example, \cite{em}), and involves 
defining the following functional:
\be
\sigma ^{2} [ \Psi ] =
\frac {\langle \Psi | {\cal {H}} |\Psi \rangle }{\langle
\Psi  |{\cal {A}} | \Psi \rangle }
=\frac {\langle {\cal {H}} \rangle }{\langle {\cal {A}} \rangle }
\ee
for any trial function $\Psi $. 
The lowest eigenvalue of the system (\ref{varsyst}) gives 
a lower bound for this functional.
Therefore,  there are negative eigenvalues $\sigma ^{2}$ 
(which correspond to unstable modes) if we can find any 
function $\Psi $ which satisfies 
$ \sigma ^{2} [ \Psi ]  <  0 $, with 
$\langle  {\cal {A}}  \rangle  <  \infty $.
The advantage of this approach is that it is easier to find 
trial functions which satisfy these criteria than it is to 
find eigenfunctions, which often involves numerical analysis.
The disadvantage is that we do not obtain precise information
about the number or magnitude of the negative eigenvalues. 
In this subsection we are interested in showing that the presence 
of the dilaton and Gauss-Bonnet term in our model is not 
sufficient to render the gauge field hair topologically stable.
We shall return to the question of the number of unstable modes
in the next subsection.

The inner product of two trial functions is defined as follows:
\be
\langle \Psi | \Phi \rangle 
=\int _{r_{h}}^{\infty } e^{\phi } e^{-(\Gamma -\Lambda )/2}
{\bar {\Psi }}\Phi \,  dr ,
\label{innprod}
\ee
and with respect to this inner product the operator ${\cal {H}}$ is 
self-adjoint, whilst ${\cal {A}}$ is positive definite.
This inner product (\ref{innprod})
is slightly different from the one usually employed 
(see, for example, \cite{em}) due to the $e^{\phi }$ term.  
This term is crucial if ${\cal {H}}$ is to be self-adjoint.
We consider the following trial functions:
\ba
\delta b & = & -w' Z_{k}(r) 
\nonumber \\
\delta \nu & = & (w^{2}-1)  Z_{k}(r),
\ea
which are the same as those used in the Einstein-Yang-Mills-Higgs
case \cite{em} 
(these are appropriate due to our inner product (\ref{innprod}), 
which effectively absorbs all the $\phi $ 
dependence in this sector).
In order to define the functions $Z_{k}$, first introduce a 
new co-ordinate $\rho $:
\be
\frac {d\rho }{dr}=e^{\phi } e^{-(\Gamma -\Lambda )/2}.
\label{ntortoise}
\ee
This is not the usual ``tortoise'' co-ordinate because of the $\phi $
dependence.
Then we define the sequence of functions $Z_{k}(\rho )$ by
\cite{volkov}
\be
Z_{k}(\rho )=Z\left( \frac {\rho }{k} \right);
\qquad
k=1,2,\ldots 
\ee
where $Z(\rho )$ is an even function which is equal to unity
for $\rho \in [0,C]$, vanishes for $\rho >C+1$, and satisfies
\be
-D\le \frac {dZ}{d\rho } <0 \qquad
{\mbox {for $\rho \in [C,C+1] $}},
\ee
with $C$ and $D$ arbitrary positive constants.

With these trial functions, after a lengthy calculation, we have
\ba
\langle  {\cal {A}}  \rangle & = &
\int _{r_{h}}^{\infty } dr \, e^{\phi } e^{-(\Gamma -\Lambda )/2} 
Z_{k}^{2} \left[
r^{2} e^{-\Lambda } w'^{2} +2(w^{2}-1)^{2} \right]
\nonumber
\\
\langle  {\cal {H}}  \rangle & = & 
-\int _{r_{h}}^{\infty } dr \, 2 e^{\phi } e^{(\Gamma -\Lambda )/2}
{\cal {J}}
\nonumber \\
& &
+\int _{r_{h}}^{\infty } dr \, 2 e^{\phi } e^{(\Gamma -\Lambda )/2}
{\cal {J}}(1-Z_{k}^{2}) 
\nonumber \\
& & 
+\int _{r_{h}}^{\infty } dr \, 2 e^{\phi } e^{(\Gamma -\Lambda )/2}
(w^{2}-1)^{2} Z_{k}^{'2} ,
\label{hexp}
\ea
where
\be
{\cal {J}}=w'^{2} +\frac {e^{\Lambda }}{r^{2}} (w^{2}-1)^{2}
\ge 0,
\ee
and all boundary terms have vanished due to the definition 
of the functions $Z_{k}$.
Immediately it can be seen that the expectation values of 
${\cal {A}}$ and ${\cal {H}}$ are finite for each value of $k$, 
and the expectation value of
${\cal {A}}$ is positive, as expected. 
The second and third terms in (\ref{hexp}) converge to zero as
$k\rightarrow \infty $, and hence choosing $k$ sufficiently large, 
we obtain a negative expectation value for ${\cal {H}}$.
We conclude that the system of perturbation equations (\ref{varsyst})
has at least one negative mode, and hence that the black holes are 
unstable in the sphaleronic sector.

The other equation in this sector (\ref{2}) is known as the 
{\em {Gauss constraint}}.
However, we do not need to consider it here, since it
is automatically satisfied by eigenfunctions of the system
(\ref{varsyst}).
To see this, consider the ``pure gauge'' functions given by
\be
\delta b = {\cal {F}}', \qquad
\delta \nu = -w {\cal {F}}
\label{gaugemodes}
\ee
where ${\cal {F}}$ is an arbitrary (differentiable) function of $r$.
The perturbations (\ref{gaugemodes}) satisfy (\ref{varsyst}) with
eigenvalue $\sigma ^{2}=0$ for any ${\cal {F}}$.
Therefore, if $(\delta b, \delta \nu )$ are eigenfunctions with
eigenvalue $\sigma ^{2}\neq 0$, they must be orthogonal to the
``pure gauge'' modes: 
\be
\left\langle \left(
\begin{array}{c} {\cal {F}}' \\ -w{\cal {F}} 
\end{array} \right) \right| {\cal {A}} \left| \left(
\begin{array}{c} \delta b \\ \delta \nu 
\end{array} \right) \right\rangle =0.
\ee
Performing an integration by parts, we obtain
\be
\int _{r_{h}}^{\infty } dr \, e^{\phi } 
e^{-\frac {1}{2} (\Gamma +\Lambda )} r^{2} 
{\cal {F}} 
{\cal {G}}
=0,
\ee
where ${\cal {G}}$ is the expression in brackets in the Gauss
constraint (\ref{2}).
Since ${\cal {F}}$ is arbitrary, and the integrand is assumed to be
continuous, the Gauss constraint must be satisfied by the
perturbations $\delta b$ and $\delta \nu $.

We note that the analysis of this subsection
applies equally well to the 
coloured black holes
of \cite{kt} as well as the magnetically charged black holes 
which occur when $w\equiv 0$ \cite{torii}.
The instability of the magnetically charged black holes
may be understood by considering them as coloured black holes in the
limit in which the number of nodes of the gauge field function $w$ 
goes to infinity.
Therefore, they correspond to saddle points of the action
(\ref{action}) due to the fact that the gauge group is non-Abelian.
In addition, since the sphaleronic sector perturbation equations do 
not depend on the Gauss-Bonnet term, and the proof of instability 
depends not on the details of the equilibrium solutions, 
but only on their global features (such as
the existence of the event horizon) the analysis also applies to the 
EYMD black holes \cite{lavr}.
We emphasize that the calculations necessary to obtain the 
expectation value (\ref{hexp}) made use only of the static 
equilibrium equation for $w$ (which has the same form for 
both the Gauss-Bonnet and EYMD models) and not that for $\phi $, 
which does involve the Gauss-Bonnet term.
Therefore the EYMD black holes are also unstable 
in the sphaleronic sector.
This is what would be anticipated, but 
previous studies of the stability 
of the EYMD system have concentrated only on the gravitational sector
\cite{lavr}.
Therefore the presence of the dilaton (either with or 
without an associated Gauss-Bonnet term) is not sufficient 
to remove the topological instabilities of the Yang-Mills field.  
So far, the only way to do this is to introduce a negative 
cosmological constant \cite{ew}.

\subsection{Counting the number of unstable modes}

Having shown that there exist instabilities in the sphaleronic
sector, we now count the number of unstable modes.
This subsection extends the method of \cite{volcount}, which 
counts the number of sphaleronic instabilities of Einstein-Yang-Mills
(EYM) black holes, 
to systems involving a dilaton field and Gauss-Bonnet curvature
term.

Firstly, we define the usual ``tortoise'' co-ordinate $r^{*}$
(compare $\rho $ (\ref{ntortoise})):
\be
\frac{dr^{*}}{dr}=e^{-(\Gamma-\Lambda)/2}
\label{tortoise}
\ee
and introduce the quantities
\be
\chi=\frac{r^2}{2}\,e^{-(\Gamma+\Lambda)/2} e^\phi \, \delta b\,,
\qquad \gamma^2=\frac{2e^\Gamma}{r^2}\,.
\label{eq3b}
\ee
Then the perturbed equations for the sphaleronic sector 
(\ref{varsyst}) can be written in the more compact form
\ba
0 & = & 
\frac {d\chi }{dr^{*}} + e^\phi w\,\delta\nu 
\nonumber
\\
\sigma^2 \chi & = & 
w^2 \gamma^2 \chi + e^\phi 
\left( w\,\frac {d}{dr^{*}}\delta\nu  - 
\frac {dw}{dr^{*}} \delta \nu \right) 
\label{eq3c}
\\
-\sigma^2 e^\phi w\,\delta\nu & = &
\frac {d}{dr^{*}}
\left[ w^2 \gamma^2 \chi + e^\phi
\left( w\, \frac {d}{dr^{*}}\delta \nu  -
\frac {dw}{dr^{*}} \delta\nu \right) \right] . \nonumber
\ea
The first of these equations is the Gauss constraint 
which we assume to hold even in the case that $\sigma=0$.
In that case, we can easily prove that the third equation
is a direct consequence of the first two and it will be
ignored hereafter.
In terms of the new function 
\be
u=\frac {e^{-\frac {\phi }{2}}\chi }{w},
\ee 
the first two perturbation equations in (\ref{eq3c}) can be
rearranged to give
\be
-\frac {d^{2}u}{dr^{*2}} 
+{\cal {U}}_{1}(r^{*}) u
=\sigma^2 u\,,
\label{diff2}
\ee
where
\be
{\cal {U}}_{1}(r^{*})
=  
\frac{1}{2}\,\gamma^2(1+w^2) +
\frac{2}{w^{2}} \left( \frac {dw}{dr^{*}} \right) ^{2}
+ \frac{2}{w}\frac {dw}{dr^{*}} \frac {d\phi }{dr^{*}}
+\frac{1}{4} \left( \frac {d\phi }{dr^{*}} \right) ^{2}
-\frac {1}{2} \frac {d^{2}\phi }{dr^{*2}} .
\ee
This is a standard Schr\"odinger equation, but the potential
${\cal {U}}_{1}$ is not regular because $w$ has zeros.
Therefore we need to use the method of \cite{volcount} to map
this equation to a ``dual'' Schr\"odinger equation which
will have a regular potential.

The ``pure gauge'' modes (\ref{gaugemodes}) give a solution
of (\ref{diff2}) when $\sigma =0$, namely
\be
u_{0}=\frac {e^{\phi /2}}{w\gamma ^{2}}
\frac {d{\cal {F}}}{dr^{*}} ,
\ee
where, in order that the Gauss constraint is satisfied (even
though $\sigma =0$) the arbitrary function ${\cal {F}}$
must satisfy the differential equation
\be
\frac {d}{dr^{*}} \left( \frac {e^{\phi }}{\gamma ^{2}}
\frac {d\cal {F}}{dr^{*}} \right) 
=e^{\phi }w^{2} {\cal {F}} .
\label{diffF}
\ee
The function $u_{0}$ can be used to factorize the operator 
acting on $u$ in (\ref{diff2}):
\be
Q^{+}Q^{-} 
=  - \left( \frac{d}{dr^{*}}
+\frac{1}{u_0} \frac {du_{0}}{dr^{*}}\right)
\left(\frac{d}{dr^{*}} 
-\frac{1}{u_0} \frac {du_{0}}{dr^{*}} \right) 
=  
-\frac{d^2}{dr^{*2}}+ \frac{1}{u_0} \frac {d^{2}u_{0}}{dr^{*2}} .
\ee
By defining $\psi=Q^-u$ and applying 
$Q^-$ again, we obtain the ``dual'' eigenvalue equation
\be
Q^- Q^+ \psi=
-\frac {d^{2}\psi }{dr^{*2}} 
+{\cal {U}}_{2}(r^{*}) \psi=\sigma^2 \psi\, ,
\label{diff3}
\ee
where
\be
{\cal {U}}_{2}(r^{*}) 
=  
\frac{1}{2}\,\gamma^2(3w^2-1) 
+ 2\,\frac {d}{dr^{*}}(w^2 Y)
+ \frac{1}{2} \frac {d^{2}\phi }{dr^{*2}}
+\frac{1}{4}\left( \frac {d\phi }{dr^{*}} \right) ^{2}.
\ee
In the above, $Y$ is a function of $r^{*}$ defined as
\be
Y=-\gamma^2 {\cal {F}} 
\left( \frac {d{\cal {F}}}{dr^{*}} \right) ^{-1}
\label{zdef}
\ee
and satisfying the equation
\be
\frac {dY}{dr^{*}}
=-\gamma^2 + w^2 Y^2 + \frac {d\phi }{dr^{*}} Y\,.
\label{diffZ}
\ee
The equation (\ref{diff3}) is another standard Schr\"odinger equation,
but now we have a potential which is regular everywhere outside the
event horizon, provided that $Y$ is a regular, well-defined, 
bounded function for all $r^{*}$.
Assuming for the moment that this is the case (we shall return to 
this issue shortly), the solution $\psi _{0}$ of (\ref{diff3})
when $\sigma =0$ will then have the same number of zeros as there
are eigenfunctions with $\sigma ^{2}<0$, that is, the number
of unstable modes. 
The zero energy solution $\psi_0$ satisfies the equation
\be
Q^+\psi_0
= \left( -\frac{d}{dr^{*}}-\frac{1}{u_0}\frac {du_{0}}{dr^{*}}
\right) \psi_0
 = 
\left( -\frac{d}{dr^{*}}+ w^2 Y 
+\frac{1}{w}\frac {dw}{dr^{*}}
+ \frac{1}{2} \frac {d\phi }{dr^{*}}\right) \psi_0
=  0, 
\ee
leading to the solution
\be
\psi_0= w\,e^{\phi/2}\,\exp \left(
\int_{0}^{r^{*}} w^2 Y\,dr^{*} \right) \,.
\label{zero}
\ee
The function $\psi _{0}$ then has the same number of zeros
as $w$ if the last factor on the right-hand-side of (\ref{zero})
is regular for all $r^{*}$.
Assuming, in addition, that $\psi _{0}$ is normalizable,
this leads us to conclude that the number of unstable
modes in the sphaleronic sector equals the number of zeros
of the gauge field function $w$ for the coloured stringy black holes.

At this stage we need to check our assumptions, namely
that the function $Y$, 
satisfying the differential equation (\ref{diffZ}),
is regular for all $r^{*}$ and vanishes sufficiently quickly
in the asymptotic regimes $r^{*} \rightarrow \pm \infty $
so that the integral in (\ref{zero}) is finite everywhere.
We also require that the function $\psi _{0}$ given by
(\ref{zero}) is a normalizable eigenfunction.
Firstly, from the definition (\ref{zdef}) of $Y$, it can be
seen that $Y$ will be a regular function as long as ${\cal {F}}$
is regular, and $\frac {d{\cal {F}}}{dr^{*}}\neq 0$.
The behaviour of ${\cal {F}}$ is most easily found by considering
the differential equation (\ref{diffF}) in terms of the radial
co-ordinate $r$:
\be
{\cal {F}}''+\left( \phi ' +\frac {2}{r} -\frac {\Gamma '}{2}
-\frac {\Lambda '}{2} \right) {\cal {F}}'-
\frac {2e^{\Lambda }w^{2}}{r^{2}} {\cal {F}} =0.
\label{diffFr}
\ee
This equation has regular singular points at $r=r_{h},\infty $.
Near $r=r_{h}$, the standard Frobenius method reveals that
either ${\cal {F}} \sim O(1)$, or ${\cal {F}} \sim O(r-r_{h})$
as $r \rightarrow r_{h}$.
We consider the solution of (\ref{diffFr})
which has the behaviour ${\cal {F}} \sim (r-r_{h})$.
Since (\ref{diffFr}) has only one other singular point, at infinity,
this solution can be extended to a solution regular for all $r$.
Then, as $r\rightarrow \infty $,
applying the Frobenius method again shows that either
${\cal {F}} \sim O(r)$ or ${\cal {F}} \sim O(r^{-2})$.
It must be the case that
${\cal {F}}\sim r$, since we can show
that ${\cal {F}}$ cannot vanish at both $r=r_{h}$ and
as $r\rightarrow \infty $, as follows.
From (\ref{diffF}) we have
\ba
0 & \le & 
\int _{r^{*}_{0}}^{r^{*}_{1}} e^{\phi }w^{2} {\cal {F}}^{2} \, dr^{*}
\nonumber \\ 
&  =  & \int _{r^{*}_{0}}^{r^{*}_{1}} {\cal {F}} 
\frac {d}{dr^{*}} \left( \frac {e^{\phi }}{\gamma ^{2}}
\frac {d{\cal {F}}}{dr^{*}} \right) \, dr^{*}
\nonumber \\
&  = &  
\left[ \frac {{\cal {F}}e^{\phi }}{\gamma ^{2}} 
\frac {d{\cal {F}}}{dr^{*}} \right] _{r^{*}_{0}}^{r^{*}_{1}}
-\int _{r^{*}_{0}}^{r^{*}_{1}} \frac {e^{\phi }}{\gamma ^{2}} 
\left( \frac {d{\cal {F}}}{dr^{*}} \right) ^{2} \,
dr^{*}.
\ea
Taking $r^{*}_{0}\rightarrow -\infty $ (corresponding to 
$r\rightarrow r_{h}$), $r^{*}_{1} \rightarrow \infty $
($r\rightarrow \infty $),
we obtain a contradiction if ${\cal {F}}\rightarrow 0$ for 
both limits (if ${\cal {F}}$ is not identically zero).
Similarly, if $d{\cal {F}}/dr^{*}=0$ for $r^{*}=r^{*}_{1}$,
and taking $r^{*}_{0}\rightarrow -\infty $, we also obtain
a contradiction, which means that $d{\cal {F}}/dr^{*}$ cannot
be zero for $r^{*}\in (-\infty , \infty )$.
Therefore $Y$ is a regular function of $r^{*}$ for all $r^{*}$. 
In addition, the differential equation (\ref{diffZ}) shows 
that 
\ba
Y & \sim & -\frac{1}{w_h^2 r^{*}} \rightarrow 0 \quad 
{\mbox {as $r^{*} \rightarrow -\infty $}}, 
\nonumber \\ 
Y & \sim & -\frac{1}{r^{*}} \rightarrow 0 \quad 
{\mbox {as $r^{*} \rightarrow \infty $.}}
\ea
This means that $|\psi _{0}|^{2}\sim (r^{*})^{-2}$
as $r^{*}\rightarrow \pm \infty $, and so $\psi _{0}$ is
a normalizable wave function.
Therefore the conditions necessary for our conclusion, that there
are as many negative modes as zeros of the function $w$, to hold
are satisfied. 

The analysis of this subsection once again reveals the topological
nature of the instabilities, since the precise details of the
equilibrium solutions were not needed,  but only general properties
such as the behaviour in the asymptotic regions and the number
of nodes of the gauge function.
In particular, the only equilibrium field equation used in the
calculations
is that for $w$, which does not depend explicitly on the
Gauss-Bonnet term.
Therefore, the result holds for all values of $\beta $, so both the
coloured 
black holes with a Gauss-Bonnet curvature term and the EYMD solutions
are covered.
The number of unstable modes is exactly the same as for the EYM 
black holes \cite{volcount}, so the stringy corrections make 
no difference to the topological instabilities.

So far in this subsection we have been concerned with the coloured
black holes, since we have assumed implicitly that $w$ does not
vanish identically.
For the magnetically charged black holes, where $w\equiv 0$,
the sphaleronic sector perturbation equations (\ref{varsyst})
reduce to $\delta b\equiv 0$ (implying that the Gauss 
constraint is satisfied) and
\be
\sigma ^{2} \delta \nu +\frac {d^{2}}{dr^{*2}} \delta \nu
+\frac {d\phi }{dr^{*}} \frac {d}{dr^{*}} \delta \nu
+\frac {e^{\Gamma }}{r^{2}} \delta \nu =0.
\ee
This can be cast into the form of a standard Schr\"odinger equation
for the variable $\xi =e^{\phi /2}\delta \nu $:
\be
\sigma ^{2} \xi +\frac {d^{2}\xi }{dr^{*2}} -U_{S}(r^{*})\xi =0 ,
\label{xieqn}
\ee
where the potential is
\be
U_{S}(r^{*})= 
\frac {1}{2}\frac {d^{2}\phi }{dr^{*2}} 
+\frac {1}{4} \left( \frac {d\phi }{dr^{*}} \right) ^{2}
-\frac {e^{\Gamma }}{r^{2}}.
\label{usphal}
\ee
In the next section we shall find that a Schr\"odinger equation 
with this potential also governs the gravitational sector
perturbations of the magnetically charged black holes.
We shall be able to appeal to catastrophe theory to show that
this equation (\ref{xieqn}) has an infinite number of unstable
modes, so that the magnetically charged black holes have
infinitely many sphaleronic instabilities. 
This result was anticipated from regarding the magnetically
charged black holes as the limit of coloured black holes in which
the number of zeros of $w$ goes to infinity.

As we mentioned in the beginning of this section, our analysis is
also applicable to the stability analysis of the sphaleronic sector
of the neutral black holes. Although the corresponding background
solutions have no gauge field, a sphaleronic sector arises when one
applies small gauge perturbations to the system. From
equation~(\ref{zero}),
we can easily see that, in this case, the function $\psi _{0}$ has
no nodes since, by definition, $w^{2} \equiv 1$ everywhere. This
confirms the absence of unstable modes in the sphaleronic sector of
the neutral black holes and demonstrates the stability
of this family of solutions even under gauge perturbations.

\section{Gravitational sector perturbations}

We now turn to the gravitational sector perturbations.
The perturbation equations in this sector, as shall be seen below, 
are extremely unattractive (in \cite{kmrtw2} 
a great deal of complex numerical work was necessary) and so 
we shall appeal to the catastrophe theory analysis of \cite{torii}.
In\cite{torii}
the authors investigated the properties of the 
static solutions with varying horizon radius $r_{h}$,
fixing the values
of the coupling constants $\alpha '/g^{2}$ and $\beta $.
In this section we shall firstly consider the gravitational
sector of the coloured and magnetically charged black holes.
Then, we shall confirm that the addition of the perturbation
$\delta w$ in the gravitational sector of the neutral black holes
does not introduce any instabilities. Finally, we will examine the
stability of the  electrically charged black holes, which need to be
studied separately because $a_{0}\neq 0$ for the equilibrium solutions.

\subsection{Coloured and magnetically charged
black holes}

The gravitational sector perturbation equations for the
neutral, coloured and magnetically charged black holes 
using the temporal gauge take the form:
\vspace*{3mm}
\begin{mathletters}
\label{gravpert}
\ba
&~&  0 = 
\delta\phi''+\delta\phi'\, \left( \frac{\Gamma'}{2} -\frac{\Lambda'}{2}+
\frac{2}{r}\right)  - \delta\phi\, \left[ \phi'' + \phi'\,
\left( \frac{\Gamma'}{2}-\frac{\Lambda'}{2} + \frac{2}{r}
\right) \,\right] - e^{\Lambda-\Gamma} \delta \ddot{\phi} 
+\frac{\alpha'e^{\phi}}{g^2 r^2}\,(1-e^{-\Lambda})\, 
(e^{\Lambda-\Gamma} \delta \ddot{\Lambda}- \delta\Gamma'')
\nonumber \\ & & \hspace*{0.5cm}
+\delta\Gamma' \left\{ \frac{\phi'}{2}- \frac{\alpha' e^{\phi}} 
{g^2 r^2}\, \left[  \Lambda' e^{-\Lambda} +(1-e^{-\Lambda})
\left( \Gamma'-\frac{\Lambda'}{2} \right) \right] \right\}
+\frac{2\alpha' e^\phi}{g^2 r^2}\,\biggl\{\frac{e^\Lambda}{r^2}
\,(1-w^2)\,w\,\delta w-w'\delta w' \biggr\}
\nonumber \\ & & \hspace*{0.5cm}
+\delta \Lambda\, \frac{\alpha' e^{\phi}}{g^2 r^2}\, 
\left\{ \Gamma' \Lambda' e^{-\Lambda} -e^{-\Lambda} 
\left[\Gamma'' + \frac{\Gamma'}{2}\, (\Gamma'-\Lambda')\right] 
-\frac{e^\Lambda}{2r^2}\,(1-w^2)^2\right\}
- \delta\Lambda' \left\{ \frac{\phi'}{2} -
\frac{\alpha' e^{\phi}}{g^2 r^2} \frac{\Gamma'}{2} 
(1-3e^{-\Lambda } ) \right\};
\label{1}\\[3mm]
&~&  0  =  
-e^{\Lambda-\Gamma}\,\delta\ddot{w}+\delta w'' + 
\delta w'\biggl(\phi'+\frac{\Gamma'-\Lambda'}{2}\biggl) 
+w'\biggl(\delta\phi'+
\frac{\delta\Gamma'-\delta\Lambda'}{2}\biggl)
+\frac{e^\Lambda}{r^2}\,\biggl[\delta\Lambda\,w\,(1-w^2)
+\delta w\,(1-3w^2)\biggr];
\label{5}\\[3mm]
&~& 0  = 
\delta\Lambda'\left[ \,1+ \frac{\alpha' e^{\phi} \phi'}{2g^2r}\,(1-3
e^{-\Lambda})\,\right]
+\delta\phi' \left[ \,-\frac{r \phi'}{2} +\frac{\alpha' e^{\phi}}{2g^2 r}\,
\Lambda' \,(1-3e^{-\Lambda})-\frac{2\alpha'e^\phi}{g^2r}\,
\phi'\,(1-e^{-\Lambda})\,\right]
-\delta\phi''\,\frac{\alpha'e^\phi}{g^2r}\,(1-e^{-\Lambda})
\nonumber \\ & &\hspace*{0.5cm}
+ \delta\Lambda \left\{ \,\frac{e^\Lambda}{r}  + 
\frac{\alpha' e^\phi}{g^2 r}\,\left[e^{-\Lambda}\,
\frac{3 \phi'\Lambda'}{2} - e^{-\Lambda}\,(\phi''+\phi'^2)-
\frac{e^\Lambda}{4r^2}\,(1-w^2)^2\right]\right\}
+\frac{\alpha' e^{\phi}}{g^2 r}\,\biggl[\frac{e^\Lambda}{r^2}\,
(1-w^2) \,w\,\delta w -w'\delta w'\biggr]
\nonumber \\ & & \hspace*{0.5cm}
+\delta\phi\,\frac{\alpha' e^{\phi}}{g^2 r}\,\biggl[
\frac{\phi' \Lambda'}{2}\, (1-3e^{-\Lambda})  
-(1-e^{-\Lambda})\,(\phi''+\phi'^2)
-\frac{e^\Lambda}{4r^2}\,(1-w^2)^2-\frac{w'^2}{2}\biggr];
\label{6}\\[3mm]
&~&  0  =  
\delta\Gamma'\left[ \,1+ \frac{\alpha' e^{\phi} \phi'}{2g^2r}\,(1-3
e^{-\Lambda})\,\right] 
+ \delta\phi'\left[ \,-\frac{r \phi'}{2}+
\frac{\alpha' e^{\phi}}{2g^2 r}\,
\Gamma' \,(1-3e^{-\Lambda})\,\right]
-\delta\ddot{\phi}\,\frac{\alpha'e^\phi}{g^2r}\,e^{\Lambda-\Gamma}\,
(1-e^{-\Lambda}) 
\nonumber \\ & & \hspace*{0.5cm}
+\delta\phi\,\frac{\alpha' e^{\phi}}{2g^2 r}\,\biggl[\phi' \Gamma'
\, (1-3e^{-\Lambda})
+ \frac{e^\Lambda}{2r^2}\,(1-w^2)^2-w'^2\biggr]
-\frac{\alpha' e^{\phi}}{g^2 r}\,\biggl[\frac{e^\Lambda}{r^2}\,
(1-w^2) \,w\,\delta w+w'\delta w'\biggr]
\nonumber \\ & & \hspace*{0.5cm}
+ \delta\Lambda \left\{ \,-\frac{e^\Lambda}{r} +
\frac{\alpha' e^\phi}{2g^2 r}\,\left[3 \phi'\Lambda'e^{-\Lambda}
+ \frac{e^\Lambda}{2r^2}\,(1-w^2)^2\right]\right\};
\label{7}
\\[3mm]
&~&  0  =  
\delta\dot{\Lambda}\,\left[ 1+ \frac{\alpha' e^{\phi} \phi'}{2g^2r}
\,(1-3e^{-\Lambda}) \right] -\frac{r \phi'}{2}\,\delta\dot{\phi} 
- \frac{\alpha' e^\phi}{g^2 r}\,\biggl[(1-e^{-\Lambda}) \,
\left( \delta\dot{\phi} \,\phi'+ \delta\dot{\phi'} -
\delta\dot{\phi}\,\frac{\Gamma'}{2}\right)
+w'\delta\dot{w}\biggr] ;
\label{8}
\\[3mm]
&~&  0  = 
\delta\Gamma''\left( 1-\frac{\alpha' e^{\phi-\Lambda}}{g^2r}\,\phi'
\right) +\frac{\delta\Gamma'}{2}\,(\Gamma'-\Lambda') 
+(\delta\Gamma'-\delta\Lambda') \left( \frac{\Gamma'}{2}+\frac{1}{r}\right)
 +\phi ' \, \delta \phi ' -e^{\Lambda-\Gamma}\delta\ddot{\Lambda}
-\frac{\alpha' e^{\phi-\Lambda}}{g^2r} \biggl\{
(\delta\phi''+2\phi'\,\delta\phi')\Gamma'
\nonumber \\ & & \hspace*{0.5cm}
+ \delta\phi'\,\Gamma''
+\delta\Gamma'(\phi''+\phi'^2)+ \Bigl(\frac{\delta\phi'\Gamma'}{2}+
\frac{\phi' \delta\Gamma'}{2}\Bigr)\,(\Gamma'-3\Lambda')
+(\delta\phi-\delta\Lambda)\,\biggl[ \,\phi'\Gamma'' 
+ \Gamma'\,(\phi''+ \phi'^2) +\frac{\phi' \Gamma'}{2}\,
(\Gamma'-3 \Lambda')\biggr]
\nonumber \\ & & \hspace*{0.5cm}
+\frac{\phi' \Gamma'}{2}\,(\delta\Gamma'-3 \delta\Lambda')
+ e^{\Lambda-\Gamma}\,(\Lambda' \delta\ddot{\phi}-
\phi' \delta\ddot{\Lambda}) \biggr\}
-\frac{\alpha'e^{\phi+\Lambda}}{2g^2r^4}\,\biggl[
(\delta\phi+\delta\Lambda)\,(1-w^2)^2
-4(1-w^2)\,w\,\delta w\Biggr].
\label{9}
\ea
\end{mathletters}
The presence of the Gauss-Bonnet term means that 
these perturbation equations are considerably more complex 
than those of the EYMD system \cite{lavr}.
However, the gauge field does not make them particularly 
more complicated than for the neutral black holes with the 
Gauss-Bonnet term \cite{kmrtw2}.

We can immediately reduce the number of equations by one, by
integrating equation (\ref{8}) with respect to time to give
\be
0 = 
\delta\Lambda\,\left[ 1+ \frac{\alpha' e^{\phi} \phi'}{2g^2r}
\,(1-3e^{-\Lambda}) \right] -\frac{r \phi'}{2}\,\delta\phi 
-
\frac{\alpha' e^\phi}{g^2 r}\,\biggl[(1-e^{-\Lambda}) \,
\left( \phi ' \, \delta\phi + \delta\phi' -
\delta\phi\,\frac{\Gamma'}{2}\right)+ w'\delta w\biggr]
-\mu (r)
\label{10}
\ee
where $\mu(r)$ is an arbitrary function of $r$.
Differentiating the above equation with respect to $r$ 
and comparing it
with equation (\ref{6}) by using, 
at the same time, the time-independent equations of motion, 
a lengthy calculation gives the following differential 
equation for $\mu(r)$
\be
\mu '(r) +\mu (r) \left[ \frac {1}{2}\,(\Gamma '-\Lambda ' ) 
+\frac {1}{r} \right]=0 .
\ee
When integrated, this equation yields
$\mu(r)\propto e^{(\Lambda-\Gamma)/2}/r$ which goes to infinity when 
$r\rightarrow r_{h}$ in contradiction to our assumption of small,
bounded perturbations.
As a result, the only acceptable solution for the function 
$\mu(r)$ is the trivial one, $\mu\equiv 0$. 

For both the coloured and magnetically charged black holes, 
the behaviour of the static solutions as $r_{h}$
is varied is qualitatively the same.
Static configurations which solve the field equations exist 
for all $r_{h}$ above a certain value, at which point a 
naked singularity forms.
This is in accord with the result that there are no 
particle-like solutions in this theory \cite{donets}.
There is no critical point in the graph of black hole mass 
against horizon radius~\cite{torii}, and therefore catastrophe theory 
tells us that the stability of the solutions
does not alter as we vary $r_{h}$.  
This result can be considered from another perspective.

The system of equations (\ref{gravpert}) could be rewritten
as two coupled linear equations
for the perturbations $\delta \phi $ and $\delta w$
by eliminating $\delta \Lambda $ and $\delta \Gamma $.
(This would be a long and tedious calculation, see
\cite{kmrtw2} for the corresponding computation in the
neutral case.)
Assuming that the equilibrium solutions are analytic
in the parameter $r_{h}$, it would be possible to invoke
functional analysis theorems to show that the negative
eigenvalues $\sigma ^{2}$ of the system must
also be analytic in $r_{h}$.
Therefore the number of negative eigenvalues (corresponding
to the number of unstable modes) can only
change when
an eigenvalue passes through the value zero.
A perturbation changes the mass of the black hole by an amount
proportional to the eigenvalue.
Therefore a zero mode does not change the mass of the black hole.
Furthermore, a zero mode corresponds to a time-independent perturbation,
in other words, a small, static perturbation of the
equilibrium solutions.
Such a perturbation exists only if there are two
equilibrium solutions arbitrarily close together
for the same value of the black hole mass, that is,
at a critical point.
Therefore, when there is no critical point 
(as is the case for the coloured and magnetically charged
black holes), the stability does not change
as we vary the parameter $r_{h}$.
For the neutral black holes, there are two branches of
solutions \cite{torii}, the upper branch extending to
arbitrarily large $r_{h}$.
The analysis below would then apply to this upper branch,
and we shall below obtain agreement with the known result that this
branch of solutions is linearly stable \cite{kmrtw2}.
 
Since there is no upper bound on the value of $r_{h}$, 
we shall consider the perturbation equations as 
$r_{h}\rightarrow \infty $, in which case they will
take a particularly simple form. From this 
we shall be able to appeal to the catastrophe theory 
analysis to deduce the stability of the black holes, 
whatever their horizon radius.
First, introduce the following dimensionless variables:
\be
{\hat {r}}=\frac {r}{r_{h}}, \qquad
{\hat {t}} =\frac {t}{r_{h}}.
\label{dimless}
\ee
In that case, all the terms proportional to $\alpha'$ become of
$O(1/r_h^2)$ while all the other terms are of $O(1)$.
However, we cannot yet take the limit $r_h \rightarrow \infty$ since
the Gauss-Bonnet terms contain derivatives of the metric functions
that diverge at the limit $r \rightarrow r_h$. 
In order to resolve this, we use, once again, the tortoise coordinate 
transformation (\ref{tortoise}) which now connects $\hat{r}$ with
$\hat{r}^*$. Then, the perturbed equations take the form
\vspace*{5mm}
\ba
&~&  0  =  -\delta {\ddot {\phi }} + 
\frac {d^{2}}{d\hat{r}^{*2}} \delta \phi  
+\frac {2}{{\hat {r}}}\,e^{(\Gamma-\Lambda)/2}\frac {d}{d\hat{r}^{*}}
\delta \phi
- \left[ \frac {d^{2}\phi }{d\hat{r}^{*2}}  +  \frac {2}{{\hat {r}}}\,
e^{(\Gamma-\Lambda)/2} \frac {d\phi }{d\hat{r}^{*}} 
\right] \delta \phi  +\frac {1}{2} \frac {d\phi }{d\hat{r}^{*}}
\left( \frac {d}{d\hat{r}^{*}}\delta \Gamma  
-\frac {d}{d\hat{r}^{*}}\delta \Lambda \right) +
O\left( \frac {1}{r_{h}^{2}}  \right);
\nonumber
\\[3mm]
&~& 0  = 
-\delta {\ddot {w}} +\frac {d}{d\hat{r}^{*2}}\delta w  
+\frac {d\phi }{d\hat{r}^{*}}\frac {d}{d\hat{r}^{*}}\delta w 
+\left[ \frac {d}{d\hat{r}^{*}}\delta \phi  +\frac {1}{2}
\left( \frac {d}{d\hat{r}^{*}}\delta \Gamma  
- \frac {d}{d\hat{r}^{*}}\delta \Lambda \right)
\right] \frac {dw}{d\hat{r}^{*}} 
+\frac {e^{\Gamma}}{{\hat {r}}^{2}} \left[
w(1-w^{2}) \delta \Lambda +(1-3w^{2}) \delta w \right] ; 
\nonumber
\\[3mm]
&~&  0  =  
e^{(\Gamma-\Lambda)/2} \frac {d}{d\hat{r}^{*}} \delta \Gamma   
-\frac {{\hat {r}}}{2} \frac {d\phi }{d\hat{r}^{*}}
\frac {d}{d\hat{r}^{*}}\delta \phi 
-\frac {e^{\Gamma}}{{\hat {r}}}\,\delta \Lambda 
+ O\left( \frac {1}{r_{h}^{2}} \right) ;
\nonumber
\\[3mm]
&~& 0 =  
\delta \Lambda 
-\frac {{\hat {r}}}{2} \frac {d\phi }{d\hat{r}^{*}} \,\delta \phi 
+ O\left( \frac {1}{r_{h}^{2}} \right). 
\label{grav}
\ea
As $r_{h}\rightarrow \infty $, in accordance with the no-hair
theorem, the geometry becomes that
of a Schwarzschild black hole with 
$\phi \equiv {\mbox {const}}$ and
the Yang-Mills field superimposed on this background.
As a result, both of the
perturbations $\delta\Lambda$ and $\delta\Gamma $ vanish while the
equation for $\delta\phi$ reduces to the following form
\be
-\frac {d^{2}\lambda }{d\hat{r}^{*2}}  
+ \frac{e^{(\Gamma-\Lambda)/2}} {\hat{r}}
\left( \frac {d\Gamma }{d\hat{r}^{*}}-
\frac {d\Lambda }{d\hat{r}^{*}} \right) \lambda = \sigma^2 \lambda \, ,
\label{gravphi}
\ee
where
\be
\lambda = \delta \phi 
\exp \left( \int_{-\infty}^{\hat{r}^{*}}
\frac{e^{(\Gamma-\Lambda)/2}}{\hat{r}}\,d\hat{r}^{*} \right)
\ee
and we have considered periodic perturbations.
The above equation takes the form of a Schr\"odinger-like differential
equation with a potential which is everywhere 
regular as well as positive and vanishes as 
$\hat{r}^{*} \rightarrow \pm \infty$. 
We, thus, may conclude that the sub-sector of the dilaton 
and metric perturbations is still characterized
by the absence of any unstable modes.
Now we turn to the perturbation equation of the Yang-Mills function. 
By implementing the results derived from the above analysis, this takes
the form
\be
-\frac {d}{d\hat{r}^{*2}} \delta w
- \frac{e^\Gamma}{\hat{r}^2}\,(1-3 w^2)\,\delta w=
\sigma^2 \delta w\,.
\label{grav7}
\ee
This equation is exactly the same, to leading order in 
$1/r_{h}^{2}$, as the perturbation equation for the 
gauge field in Einstein-Yang-Mills theory without a dilaton 
or Gauss-Bonnet term.
This was to be expected since  the EYM equations also 
decouple in the limit $r_{h}\rightarrow \infty $ 
to give a Schwarzschild geometry with a gauge field on this
background.
It is known that the EYM system possesses instabilities in 
this sector (see, for example, \cite{strau}). The above 
equation for $\delta w$ is in agreement with this result.
As the background gauge function $w$ oscillates around its
zero value \cite{kt}, a potential well is formed in the
region where $w^2<1/3$,
thus, leading to the existence of bound states in the
corresponding Schr\"odinger equation.
We therefore conclude that the stringy coloured black holes
are also unstable and their instabilities are, once again, associated
with the existence of nodes of the background gauge function $w$.
It is worth noting that, in the case of coloured black holes arising in 
the presence of a negative cosmological constant in the theory,
solutions with zero number of nodes do exist and they were proven to
be linearly stable in both perturbation sectors \cite{ew}.

That the magnetically charged black holes also possess
unstable modes in this sector may be deduced, as
in the sphaleronic sector, from the fact that they are
the limit of coloured black holes in which the number
of zeros of the gauge field goes to infinity.
For the magnetically charged black holes, which follow
if we set $w=0$, the 
potential in the Schr\"odinger equation (\ref{grav7})
is everywhere negative and goes to zero as
$\hat{r}^{*}\rightarrow \pm \infty $:
\be
U(\hat{r}^{*})=-\frac {e^{\Gamma }}{{\hat r}^2} .
\ee
The standard estimate for the number of bound states 
$\frac {1}{\pi } \int _{-\infty }^{\infty }
{\sqrt {-U(\hat{r}^{*})}} \, d\hat{r}^{*} $
\cite{messiah}
then shows that there are an infinite number
of unstable modes in this case.

This analysis for very large horizon radius has confirmed what 
might have been anticipated, namely that the instabilities 
are due to the presence of the gauge field.
In addition, it is known that the EYMD black holes possess gravitational
instabilities \cite{lavr}, which is in agreement with our results,
assuming that stability does not change as $\beta $ varies.
This is a reasonable assumption, since the behaviour of the
field equations as $r_{h}\rightarrow \infty $ is the same for
all $\beta $.

As explained earlier in this subsection, catastrophe theory
tells us that the stability of the black holes does not change
as we vary $r_{h}$.
The results above, derived in the limit $r_{h}\rightarrow \infty $,
can therefore be extended to arbitrary $r_{h}$ for which black hole
solutions exist.
In particular, the coloured black holes will be unstable in this
sector, and the magnetically charged black holes will have
an infinite number of unstable gravitational modes.

We now exploit the result that the magnetically char\-ged black holes
have infinitely many unstable modes in the gravitational sector,
irrespective of the value of the horizon radius, to infer that they
also have infinitely many unstable modes in the sphaleronic sector.
For the magnetically charged black holes, and any value of the horizon
radius $r_{h}$, the perturbation 
equation for $\delta w$ (\ref{5})
decouples from the other equations in the 
gravitational sector and has the simplified form
\be
\sigma ^{2} \xi +\frac {d^{2}\xi }{dr^{*2}} -U_{G}(r^{*})\xi =0
\label{xigeqn}
\ee
where $\xi =e^{\phi /2}\delta w$ and
\be
U_{G}(r^{*})=\frac {1}{2} \frac {d^{2}\phi }{dr^{*2}}
+\frac {1}{4}\left( \frac {d\phi }{dr^{*}} \right) ^{2}
-\frac {e^{\Gamma }}{r^{2}} .
\ee
This potential is exactly the same as that
arising in the sphaleronic sector perturbations
of the magnetically charged black holes (\ref{usphal}).
Therefore, since the Schr\"odinger equation (\ref{xigeqn})
has infinitely many negative eigenvalues,
the magnetically charged black holes also have 
infinitely many unstable modes in the sphaleronic sector, 
as asserted in the previous section. 

For the neutral black holes, in the limit of large horizon 
radius, the potential in the
perturbation equation for $\delta w$ (\ref{grav7}),
after setting $w=\pm 1$, reduces to 
\be
U(\hat{r}^{*})=\frac {2e^{\Gamma }}{{\hat {r}}^{2}}.
\ee
Since this potential is everywhere positive and tends to zero
in the asymptotic regions, we can conclude that there
are no unstable modes. 
Therefore the neutral black holes are stable as we vary $r_{h}$
(we emphasize that this applies only to the upper branch
of solutions in \cite{torii}, namely the branch of solutions
which extends to arbitrarily large horizon radius and which
includes the bulk of the background black hole solutions \cite{extra}).
In the next subsection we shall show, by another method,
that the equation for $\delta w$ and general $r_{h}$
has no unstable modes.
This result will be useful in studying the stability
of the electrically charged black holes, to which we now turn.

\subsection{Electrically charged black holes}

We now turn to the electrically charged black holes, which need to be 
considered separately from the other cases because 
$a_{0}\neq 0$ for the equilibrium solutions.
As in the previous subsection, we first
consider the linearized perturbation equations involving
the metric perturbation $\delta \Lambda $:
\vspace*{4mm}
\begin{mathletters}
\label{lambdaelec}
\ba
&~& 0 = 
\delta\Lambda'\left[ \,1+ \frac{\alpha' e^{\phi} \phi'}{2g^2r}\,(1-3
e^{-\Lambda})\,\right]
+\delta\phi'
\left[ \,-\frac{r \phi'}{2}+\frac{\alpha' e^{\phi}}{2g^2 r}\,
\Lambda' \,(1-3e^{-\Lambda})-\frac{2\alpha'e^\phi}{g^2r}\,
\phi'\,(1-e^{-\Lambda})\,\right]
\nonumber \\ & & \hspace*{0.5cm} 
+ \delta\Lambda \left\{ \,\frac{e^\Lambda}{r} +
\frac{\alpha' e^\phi}{g^2 r}\,\left[e^{-\Lambda}\,\frac{3 \phi'\Lambda'}{2}
- e^{-\Lambda}\,(\phi''+\phi'^2) \right]\right\} 
-\frac {\alpha ' r}{4g^{2}} e^{\phi } e^{-\Gamma } \left[
a_{0}'^{2} \,(\delta \phi -\delta \Gamma) 
+2a_{0}' \,(\delta a_{0}'- \delta {\dot {b}}) \right]
\nonumber \\ & & \hspace*{0.5cm} 
-\delta\phi''\,\frac{\alpha'e^\phi}{g^2r}\,(1-e^{-\Lambda})
+\delta\phi\,\frac{\alpha' e^{\phi}}{g^2 r}\,\biggl[
\frac{\phi' \Lambda'}
{2}\, (1-3e^{-\Lambda})-(1-e^{-\Lambda})\,(\phi''+\phi'^2) \biggr] ;
\label{six}\\[3mm]
&~& 0  =  
\delta\dot{\Lambda}\,\left[ 1+ \frac{\alpha' e^{\phi} \phi'}{2g^2r}
\,(1-3e^{-\Lambda}) \right] - \frac{r \phi'}{2}\,\delta\dot{\phi} 
-
\frac{\alpha' e^\phi}{g^2 r}\,(1-e^{-\Lambda}) \,
\left( \delta\dot{\phi} \,\phi'+ \delta\dot{\phi'} -
\delta\dot{\phi}\,\frac{\Gamma'}{2}\right) .
\label{eight}
\ea
\end{mathletters}
Integrating (\ref{eight}) gives, as in the previous subsection,
a function $\mu (r)$ such that
\be
\mu (r) =  \delta \Lambda \left[
1+\frac {\alpha ' }{2g^{2}r} e^{\phi } \phi ' 
(1-3e^{-\Lambda }) \right]
-\frac {r\phi '}{2} \delta \phi 
-\frac {\alpha 'e^{\phi }}{g^{2}r} (1-e^{-\Lambda }) \left(
\phi ' \, \delta \phi +\delta \phi ' -\frac {\Gamma '}{2} \delta \phi 
\right) ,
\label{deltalambda}
\ee
which satisfies the following differential equation:
\be
\mu '(r) +\mu (r) \left[ \frac {1}{2} (\Gamma '-\Lambda ' ) 
+\frac {1}{r} \right]
=\frac {\alpha 'r}{2g^{2}} e^{\phi }e^{-\Gamma } a_{0}' 
\left[a_{0}' \, \delta \phi
-\delta {\dot {b}} +\delta a_{0}' 
-\frac {1}{2} a_{0}' (\delta \Gamma +\delta \Lambda ) \right] .
\label{muelec}
\ee
Since the left-hand-side of equation (\ref{muelec}) depends
only on $r$ and not on $t$, the same must be true of the
combination of perturbations in the brackets on the 
right-hand-side.
For periodic perturbations, it follows that
\be
\delta {\dot {b}} -\delta a_{0}' -a_{0}' \, \delta \phi 
+\frac {1}{2} a_{0}' (\delta \Gamma +\delta \Lambda ) =0,
\label{gaugepick}
\ee
in which case, exactly as for the coloured and 
magnetically charged black holes, $\mu (r) \equiv 0$.
It should be stressed at this point that equation (\ref{gaugepick})
does not in fact correspond to a choice of gauge.
According to the residual gauge freedom of the gauge potential
(\ref{gaugefree}) the quantity $\delta {\dot {b}}-\delta a_{0}'$
is invariant, so we cannot choose its form.
Equation (\ref{gaugepick}) therefore represents a constraint on
the gauge potential perturbations, necessary in order that the
equations (\ref{lambdaelec}) are consistent.
When $a_{0}\equiv 0$ (for the neutral, coloured and magnetically
charged black holes) the two equations for $\delta \Lambda $
are automatically consistent, regardless
of the behaviour of $w$.
The remaining gauge freedom could be used
to set $\delta a_{0}\equiv 0$ in the electrically charged case,
as for the other types of black hole solution.

For the construction of the Yang-Mills perturbation equations,
we first define the quantity $c\equiv a_{0}w$.
For the equilibrium black holes, $c$ is set to vanish identically,
although $w\equiv \pm 1$ and $a_{0}\neq 0$ \cite{torii}.
Therefore, we regard $c$ as a separate entity, with perturbation
$\delta c$, which is independent of $\delta w$ and $\delta a_{0}$.
The Yang-Mills perturbation equations then take the following form:
\begin{mathletters}
\label{YMelec}
\ba
&~& 0 =  
\pm 2e^{\phi } e^{(\Lambda -\Gamma )/2} 
\left( \delta c +\delta {\dot {\nu }} \right)
+\partial _{r} \left\{
e^{-(\Lambda +\Gamma )/2} r^{2} e^{\phi } \left[
\delta {\dot {b}} 
-\delta a_{0}'-a_{0}' \, \delta \phi
+\frac {1}{2} a_{0}' \left( \delta \Gamma +\delta \Lambda \right)
\right] \right\} ;
\\[3mm]
&~& 0  =  
\pm 2e^{\phi }e^{(\Gamma -\Lambda )/2} \left(
\delta \nu ' \pm \delta b \right)
+e^{-(\Gamma +\Lambda )/2} r^{2} e^{\phi } \left[
\delta {\ddot {b}} -\delta {\dot {a}}_{0}' 
-a_{0}' \, \delta {\dot {\phi }}
+\frac {1}{2} \left( \delta {\dot {\Gamma }} +\delta {\dot {\Lambda }} 
\right) \right] ;
\\[3mm]
&~&  0  =  
e^{\Lambda -\Gamma }e^{\phi } \left(
\delta {\ddot {\nu }} +\delta {\dot {c}} \right)
+\partial _{r} \left[
e^{(\Gamma -\Lambda )/2} e^{\phi } \left(
\delta \nu ' \pm \delta b \right) \right]
+e^{\phi } e^{(\Lambda -\Gamma )/2} a_{0} \left(
a_{0} \, \delta \nu -\delta {\dot {w}} \right) ;
\\[3mm]
&~&  0  =  
-e^{\Lambda -\Gamma } \delta {\ddot {w}} +
e^{\Lambda -\Gamma } a_{0} \, \delta {\dot {\nu }}
+\delta w'' 
+\left( \phi ' +\frac {\Gamma '}{2} -\frac {\Lambda '}{2}
\right) \delta w'
+e^{(\Lambda -\Gamma )/2} a_{0} \left( \delta c+
\delta {\dot {\nu }} \right) ;
\ea
\end{mathletters}
where the $\pm $ depend on whether $w=\pm 1$.
With the constraint (\ref{gaugepick}), 
the Yang-Mills perturbation equations (\ref{YMelec}) 
simplify greatly, and reduce to the following conditions:
\be
\delta c = -\delta {\dot {\nu }}, \qquad
\delta b = \mp \delta \nu ',
\qquad
a_{0}\, \delta \nu =\delta {\dot {w}},
\ee
and the equation for $\delta w$ then also has a simple form:
\be
\delta w'' +\left[ \phi ' +\frac {1}{2} (\Gamma ' -\Lambda ' ) 
\right] \delta w' 
- \frac {2e^{\Lambda }}{r^{2}} \delta w
=0.
\label{welec}
\ee
It is rather surprising that this equation has no time dependence.
In order to study the space dependence of $\delta w$, we are going
to consider the above perturbation as the zero eigenfunction of
the following eigenvalue problem
\be
\delta w'' +\left[ \phi ' +\frac {1}{2} (\Gamma ' -\Lambda ' ) 
\right] \delta w' 
- \frac {2e^{\Lambda }}{r^{2}} \delta w
=-\sigma ^{2} \delta w .
\label{wneut}
\ee
We note as an aside that (\ref{wneut}) is the equation 
satisfied by periodic perturbations $\delta w$ in the case
of neutral equilibrium black holes.
Here we shall be using (\ref{wneut}) as a tool to show
that there are no solutions of (\ref{welec}) representing
physical perturbations.
It should be emphasized that (\ref{welec}) does not restrict the
time dependence of the perturbation $\delta w$ at all, although the
spatial dependence for each time $t$ will be that of the solution
of (\ref{wneut}) when $\sigma ^{2}=0$.

In order to get (\ref{wneut}) in the form of a Schr\"odinger
equation, it is convenient to define another new co-ordinate,
${\cal {R}}$, by
\be
\frac {d{\cal {R}}}{dr}=e^{-\phi }e^{-(\Gamma -\Lambda )/2}.
\ee
Again, the presence of the $e^{-\phi }$ term means that this
is not the ``tortoise'' co-ordinate (compare (\ref{ntortoise}) and
(\ref{tortoise})).
Then (\ref{wneut}) becomes
\be
-\frac {d^{2}}{d{\cal {R}}^{2}}\delta w +
\frac {2e^{\Gamma }e^{2\phi }}{r^{2}} \delta w
=\sigma ^{2} e^{2\phi } \delta w.
\label{wtort}
\ee
\noindent
This equation is not quite of the standard Schr\"odinger
form, due to the $e^{2\phi }$ multiplying the $\sigma ^{2}$.
However, since $e^{2\phi }>0$, the standard theorems still apply
(see, for example \cite{birkoff}), so
in particular there can be no negative eigenvalues $\sigma ^{2}$
because the potential
\be
U=\frac {2e^{\Gamma }e^{2\phi }}{r^{2}} 
\ee
goes to zero as ${\cal {R}}\rightarrow \pm \infty $ (i.e. at the 
event horizon and at infinity), and $U\ge 0$ everywhere.
Since there are no negative eigenvalues for $\sigma ^{2}$,
this means that the solution of (\ref{wtort}) with $\sigma =0$
(if it exists) can have no zeros. 
For a physical perturbation $\delta w$, we require that 
$\delta w \rightarrow 0$ as ${\cal {R}}\rightarrow \pm \infty $.
Since $\delta w$ has no zeros, it must be of one sign and have at 
least one maximum (if it is everywhere positive) or at least one
minimum (if negative).
By (\ref{wtort}), $d^{2}\delta w/d{\cal {R}}^{2}$ has the same
sign as $\delta w$, since the potential $U$ is positive.
Therefore, if $\delta w>0$, at a stationary point 
$d^{2}\delta w/d{\cal {R}}^{2}>0$ also, which means that $\delta w$
has a minimum.
This is in contradiction with the requirement that $\delta w$
tends to zero in the asymptotic regime.
A similar argument holds if $\delta w$ is everywhere negative.
The only possible solution of (\ref{welec})
is therefore $\delta w \equiv 0$, so that
\be
\delta w=\delta \nu =\delta b =\delta c =0.
\ee 
We note that the fact that there are no negative eigenvalues
$\sigma ^{2}$ of (\ref{wneut}) confirms that there are
neutral black holes with no instabilities, even when embedded
in the non-Abelian gauge group.

For the electrically charged black holes we are left, 
as with the neutral black holes, with effectively
only a gravitational sector, 
consisting of $\delta \phi $, $\delta \Gamma $, 
$\delta \Lambda $, and $\delta a_{0}$, the latter
being constrained by (\ref{gaugepick}) to be given by:
\be
\delta a_{0}' = a_{0}' \, \delta \phi 
-\frac {1}{2} a_{0}'\left( \delta \Gamma -\delta \Lambda \right) .
\label{deltaacons}
\ee
We could use the remaining gauge freedom to set
$\delta a_{0}\equiv 0$, but this will make no difference
to our analysis.
The remaining perturbation equations take the form:
\begin{mathletters}
\label{gravelec}
\ba
&~& 0 = 
\delta\phi''+\delta\phi'\, \left( \frac{\Gamma'}{2} 
-\frac{\Lambda'}{2}+
\frac{2}{r}\right) 
- \delta\phi\, \left[ \phi'' + \phi'\,
\left( \frac{\Gamma'}{2}-\frac{\Lambda'}{2} + \frac{2}{r}
\right) \,\right] - 
e^{\Lambda-\Gamma} \delta {\ddot {\phi }}
+\frac{\alpha'e^{\phi}}{g^2 r^2}\,(1-e^{-\Lambda})\, 
(e^{\Lambda-\Gamma} \delta {\ddot {\Lambda }}-\delta\Gamma'')
\nonumber\\ & & \hspace*{0.5cm}
+\delta\Gamma' \left\{ \frac{\phi'}{2}
- \frac{\alpha' e^{\phi}} 
{g^2 r^2}\, \left[  \Lambda' e^{-\Lambda} +(1-e^{-\Lambda})
\left( \Gamma'-\frac{\Lambda'}{2} \right) \right] \right\}
- \delta\Lambda' \left\{ \frac{\phi'}{2} -
\frac{\alpha' e^{\phi}}{g^2 r^2}\,\frac{\Gamma'}{2}\,( 1-3e^{-\Lambda})
\right\}
\nonumber \\ & & \hspace*{0.5cm}
+\delta \Lambda\, \frac{\alpha' e^{\phi}}{g^2 r^2}\, 
\left\{ \Gamma' \Lambda' e^{-\Lambda} -e^{-\Lambda} 
\left[\Gamma'' + \frac{\Gamma'}{2}\, (\Gamma'-\Lambda')
\right] \right\}
+\frac {\alpha '}{2g^{2}}e^{\phi } e^{-\Gamma }\left[
3a_{0}'^{2} \, \delta \phi -2a_{0}'^{2} \, \delta \Gamma
-a_{0}'^{2} \, \delta \Lambda \right] ;
\label{one}
\\[4mm]
&~& 0 =  
\delta\Gamma'\left[ \,1+ \frac{\alpha' e^{\phi} \phi'}{2g^2r}\,(1-3
e^{-\Lambda})\,\right]
+ \delta\phi'\left[ \,-\frac{r \phi'}{2}
+\frac{\alpha' e^{\phi}}{2g^2 r}\,
\Gamma' \,(1-3e^{-\Lambda})\,\right] 
-\delta\ddot{\phi}\,\frac{\alpha'e^\phi}{g^2r}\,e^{\Lambda-\Gamma}\,
(1-e^{-\Lambda})  
\nonumber \\ & & \hspace*{0.5cm} 
+ \delta\Lambda \left[ \,-\frac{e^\Lambda}{r} +
\frac{3\alpha' e^\phi}{2g^2 r}\, \phi'\Lambda'e^{-\Lambda}\right]
+\delta\phi\,\frac{\alpha' e^{\phi}}{2g^2 r}\, \phi' \Gamma'
\, (1-3e^{-\Lambda})
+\frac {\alpha ' r}{4g^{2}} e^{\phi } e^{-\Gamma } \left[
3a_{0}'^{2} \, \delta \phi 
-2 a_{0}'^{2} \, \delta \Gamma 
-a_{0}'^{2}\, \delta \Lambda \right] ;
\label{seven}\\[4mm]
&~& 0 = 
\delta\Gamma''\left( 1-\frac{\alpha' e^{\phi-\Lambda}}{g^2r}\,\phi'
\right) +\frac{\delta\Gamma'}{2}\,(\Gamma'-\Lambda') 
+(\delta\Gamma'-
\delta\Lambda') \left( \frac{\Gamma'}{2}+\frac{1}{r}\right)
-e^{\Lambda-\Gamma}\delta\ddot{\Lambda}
-\frac{\alpha' e^{\phi-\Lambda}}{g^2r} \biggl\{
(\delta\phi''+2\phi'\,\delta\phi')\Gamma'
\nonumber \\ & &  \hspace*{0.5cm}
+\delta\phi'\,\Gamma''+\delta\Gamma'(\phi''+\phi'^2)
+ \Bigl(\frac{\delta\phi'\Gamma'}{2}
+\frac{\phi' \delta\Gamma'}{2}\Bigr)
\,(\Gamma'-3\Lambda') 
+(\delta\phi-\delta\Lambda)\,\biggl[ \,\phi'\Gamma'' +
\Gamma'\,(\phi''+ \phi'^2)   
+\frac{\phi' \Gamma'}{2}\,(\Gamma'-3 \Lambda')\biggr]
\nonumber \\ & & \hspace*{0.5cm}
+\frac{\phi' \Gamma'}{2}\,(\delta\Gamma'-3 \delta\Lambda')
+ e^{\Lambda-\Gamma}\,(\Lambda' \delta\ddot{\phi}-
\phi' \delta\ddot{\Lambda}) \biggr\} 
+\phi'\,\delta\phi'
-\frac {\alpha '}{2g^{2}} e^{\phi }e^{-\Gamma }
\left[ 3a_{0}'^{2} \, \delta \phi -2a_{0}'^{2} \, \delta \Gamma 
-2a_{0}'^{2} \, \delta \Lambda \right] ;
\label{nine}
\ea
\end{mathletters}
\noindent
where $\delta \Lambda $ is given by (\ref{deltalambda}).
This could be reduced to a single equation by using (\ref{deltalambda})
and (\ref{deltaacons}).
However, since we shall be appealing to catastrophe theory, this
does not afford any advantage in the analysis.

Now that the system has been reduced to a set of coupled equations
(\ref{gravelec}) involving the perturbations 
of the dilaton field and metric functions, we can once again 
appeal to the catastrophe theory
analysis of \cite{torii}.
They find two branches of electrically charged black hole solutions,
one of which extends out to infinite horizon radius.
We shall focus on this branch of solutions 
since it is this branch (if any) which can be stable.
As in the previous section, we consider the perturbation
equations in the limit $r_{h}\rightarrow \infty $, and
work to leading order in $1/r_{h}^{2}$.
Using the same tortoise co-ordinate $\hat{r}^*$ as in the previous
subsection, we find the equations:
\ba
&~&  0  =  -\delta {\ddot {\phi }} + 
\frac {d^{2}}{d\hat{r}^{*2}} \delta \phi  
+\frac {2}{{\hat {r}}}\,e^{(\Gamma-\Lambda)/2}\frac {d}{d\hat{r}^{*}}
\delta \phi
- \left[ \frac {d^{2}\phi }{d\hat{r}^{*2}}  +  \frac {2}{{\hat {r}}}\,
e^{(\Gamma-\Lambda)/2} \frac {d\phi }{d\hat{r}^{*}} 
\right] \delta \phi  +\frac {1}{2} \frac {d\phi }{d\hat{r}^{*}}
\left( \frac {d}{d\hat{r}^{*}}\delta \Gamma  
-\frac {d}{d\hat{r}^{*}}\delta \Lambda \right) +
O\left( \frac {1}{r_{h}^{2}}  \right);
\nonumber
\\[2mm]
&~& 0  =  
e^{(\Gamma-\Lambda)/2} \frac {d}{d\hat{r}^{*}} \delta \Gamma   
-\frac {{\hat {r}}}{2} \frac {d\phi }{d\hat{r}^{*}}
\frac {d}{d\hat{r}^{*}}\delta \phi 
-\frac {e^{\Gamma}}{{\hat {r}}}\,\delta \Lambda 
+ O\left( \frac {1}{r_{h}^{2}} \right) ;
\nonumber
\\[2mm]
&~&  0 =  
\delta \Lambda 
-\frac {{\hat {r}}}{2} \frac {d\phi }{d\hat{r}^{*}} \,\delta \phi 
+ O\left( \frac {1}{r_{h}^{2}} \right). 
\label{biggravelec}
\ea

These are exactly the same equations (without the equation
for $\delta w$) as those obtained for the other
types of black hole solutions (\ref{grav}). 
In this case, for $r_{h}\gg 1$, 
the geometry reduces to a Reissner-Nordstr\"om
black hole, with $\phi \equiv {\mbox {const}}$.
As in the Schwarzschild case, the perturbations
$\delta \Gamma $ and $\delta \Lambda $ vanish identically,
and the equation for $\delta \phi$ reduces to
(\ref{gravphi}).
The same argument then shows that there are no instabilities
in this sector for $r_{h}\gg 1$.
The catastrophe theory analysis then tells us that the 
upper branch of electrically charged black holes is stable 
for all $r_{h}$.

We conclude in this subsection that the upper branch of
electrically charged black holes is stable, whereas we
have already shown that the magnetically charged black
holes are infinitely unstable.
This may be surprising, especially since the gauge field
in both cases is essentially Abelian.
However, for the magnetically charged black holes, the
gauge potential is {\em {embedded Abelian}}, in other words,
it corresponds to the product of a $U(1)$ gauge potential
and a constant matrix.
This embedding in the non-Abelian gauge group $SU(2)$
gives rise to the infinite number of unstable modes.
As discussed earlier in this article,
the same conclusion can be reached from observing that
the magnetically charged black holes arise as the limit
of coloured black holes in which the number of nodes
of the gauge function $w$ goes to infinity.
This leads to infinitely many modes of instability
in the sphaleronic sector.
On the other hand, due to the construction of
the electrically charged black holes (in particular,
setting $c\equiv a_{0}w\equiv 0$ in the field equations
although neither $a_{0}$ nor $w$ vanish) means
that the gauge potential is genuinely $U(1)$ (without
any embedding).
Therefore the stability of these solutions is not unexpected.
In this case there is effectively no sphaleronic sector, so the other
instabilities present for the coloured and
magnetically charged black holes do not arise.

\section{Conclusions}
In this article, we have considered a generalized, string-inspired
theory of gravity that describes the non-minimal coupling of a single
scalar field, the dilaton, to gravity through the higher-derivative
Gauss-Bonnet term as well as to a non-Abelian SU(2) gauge field. This
theory has been shown in previous work \cite{kt} to admit regular and
asymptotically flat black hole solutions that are characterized
by the presence of a non-trivial dilaton and gauge field on the region
outside the horizon in contradiction with the ``no-hair" theorem of the
Theory of General Relativi\-ty. Nevertheless, the hair of these black hole
solutions is merely ``secondary" in the sense that no new charges can be
associated with the aforementioned, non-vanishing fields.
The dilatonic black holes that arise in the same framework but in the
absence of the gauge field~\cite{kmrtw1} have been already proven
\cite{kmrtw2} to be linearly stable  under small, bounded,
spacetime-dependent perturbations as they correspond to the stable branch
of the family of neutral black hole solutions.
Then, the question of the behaviour of the coloured black holes under
the same type of perturbations, and in the presence of the same stringy
corrections, naturally arises.

By making an appropriate choice of gauge, the linearized perturbation
equations, for the coloured black holes, were decoupled into two sectors,
the sphaleronic and the gravitational one. In the first sector, the
perturbation equations resembled those of the Einstein-Yang-Mills-Dilaton
theory with the explicit dependence on the Gauss-Bonnet term having been
eliminated. For the needs of our analysis, well-known methods
\cite{volcount,em}, that were previously used for the stability analysis
of black hole solutions arising in the framework of the Einstein-Yang-Mills
theory, were extended in order to accommodate the dilaton field. Then,
the existence of topological instabilities was analytically demonstrated
by making use of the method of trial functions. 
The number of unstable modes in this sector was determined by mapping 
the irregular Schr\"odinger equation of the gauge perturbations to a
``dual" regular one and it was found to be the same as
the number of zeros of the background gauge field. In the gravitational
sector, due to the complexity of the perturbation equations, continuity
arguments based on the results derived from the catastrophe theory analysis
\cite{torii} allowed us to work in the limit of infinitely large horizon
value. Although the sub-system of the metric and dilaton perturbations
was found to be stable, instabilities attributed once again to the
oscillating behaviour of the background gauge function around zero were
proven to exist. As a result, we may conclude that the accommodation of
stringy corrections to non-Abelian black holes fails to render these
solutions linearly stable.

The perturbation equations for the magnetically char\-ged and neutral black
holes, under the same type of perturbations, may easily follow from the
corresponding ones for the coloured black holes by simply choosing an
appropriate value for the background gauge function $w$. This allows us to
draw conclusions concerning the stability of these two families of
solutions
in the same framework and by using the same methods as above. Then,
the study of the sphaleronic sector reveals the existence of an infinite
number of unstable modes, for the magnetically charged black holes, and
the same holds for the gravitational sector. The above result
is in accordance with the interpretation of this family of solutions as
the limit of coloured black holes in which the number of zeros of $w$
goes to infinity. On the other hand, no unstable modes are found for
the upper branch of the neutral black holes in both sectors which
confirms the stable character of these solutions not only under metric
and scalar perturbations \cite{kmrtw2} but even under gauge-dependent
perturbations. 

Finally, the stability analysis of the electrically charged solutions
was conducted although in a different framework of perturbation equations
due to the different ansatz for the background gauge field. Nevertheless,
we were able to show that, in order for our perturbation equations to be
consistent, a constraint that involves a combination of gauge, metric and
scalar perturbations must be satisfied. In that case, we have shown that
the gauge perturbations are completely decoupled from the scalar and
metric ones, a feature which facilitates the study of each subsector. 
The gauge perturbations were found to reduce to a single equation for
$\delta w$ which, however, was shown not to accept any solutions
representing physical perturbations. On the other hand, by using the
same continuity arguments and working again in the limit of infinitely
large horizon, we proved that no unstable modes arise in the remaining
subsector of metric and scalar perturbations for the upper branch of
the electrically charged black hole solutions. In conclusion, the
bulk of the electrically charged black holes, in common with the
neutral dilatonic black holes, are stable under small,
spacetime-dependent perturbations. The stability
of these two families of solutions can be justified by their
interpretation
as the generalization of the Reissner-Nordstr\"om and Schwarzschild 
black holes, respectively, in the framework of string theory. 

\section*{Acknowledgements}
P.K. would like to acknowledge financial support by DOE grant 
DE-FG02-94ER40823 at Minnesota.
E.W. would like to thank Oriel College,  Oxford for their financial
support, and the Department of Physics, 
University of Newcastle, Newcastle-upon-Tyne, UK for hospitality.

\end{document}